\documentclass[aps, twocolumn,amsmath,amssymb,floatfix,superscriptaddress]{revtex4-1}
\usepackage[table]{xcolor}
\usepackage{graphicx}
\usepackage{amsmath,fullpage,amssymb}
\usepackage{graphics,epsfig}
\usepackage{makeidx}

\usepackage{titlesec} 
\titleformat{\subsection}[runin]
{\normalfont\bfseries}{\thesubsection{.}}{1em}{}[.]

\titleformat{\section}
{\normalfont\sffamily\large\bfseries}{\thesection{.}}{1em}{\MakeUppercase}

\usepackage[table]{xcolor}
\usepackage{graphicx}
\usepackage{amsmath,fullpage,amssymb}
\usepackage{graphics,epsfig}
\usepackage{makeidx}
\usepackage{gensymb}
\usepackage{siunitx}

\newcommand*{\citen}[1]{
  \begingroup
    \romannumeral-`\x 
    \setcitestyle{numbers}%
    \cite{#1}%
  \endgroup   
}

\def\ln{{\operatorname{ln}}}

\def\rmd{{\mathrm{d}}}

\def\rme{{\mathrm{e}}}

\def\Eq{eq}
\def\Eqs{eqs}

\def\Fig{Figure}

\def\SItext{{\em Supporting Information}}
\def\ie{{\em i.e.}}

\newcommand{\kB}{k_\textrm{B}}
\newcommand{\chgA}[1]{\textcolor{black} {#1}}

\newcommand{\trm}[1]{{\textrm{#1}}}

\usepackage{soul}	

\begin{document}


\title{Transfer Free Energies and Partitioning of Small Molecules in Collapsed PNIPAM Polymers}


\author{Matej Kandu\v{c}}
\email{matej.kanduc@ijs.si}
\affiliation{\rm\small Research Group for Simulations of Energy Materials, Helmholtz-Zentrum Berlin f\"ur Materialien und Energie, Hahn-Meitner-Platz 1, D-14109 Berlin, Germany}
\affiliation{\rm\small Jo\v{z}ef Stefan Institute, Jamova 39, SI-1001 Ljubljana, Slovenia}

\author{Won Kyu Kim}
\affiliation{\rm\small Research Group for Simulations of Energy Materials, Helmholtz-Zentrum Berlin f\"ur Materialien und Energie, Hahn-Meitner-Platz 1, D-14109 Berlin, Germany}

\author{Rafael Roa}
\affiliation{\rm\small Departamento de F\'{i}sica Aplicada I, Facultad de Ciencias, Universidad de M\'{a}laga, Campus de Teatinos s/n, E-29071 M\'{a}laga, Spain}

\author{Joachim Dzubiella}
\email{joachim.dzubiella@physik.uni-freiburg.de}
\affiliation{\rm\small Applied Theoretical Physics -- Computational Physics, Physikalisches Institut, Albert-Ludwigs-Universit\"at Freiburg, Hermann-Herder Strasse 3, D-79104 Freiburg, Germany}
\affiliation{\rm\small Research Group for Simulations of Energy Materials, Helmholtz-Zentrum Berlin f\"ur Materialien und Energie, Hahn-Meitner-Platz 1, D-14109 Berlin, Germany}


\pagenumbering{arabic}
\noindent

\parindent=0cm
\setlength\arraycolsep{2pt}

\begin{abstract}
A central quantity in the design of functional hydrogels used as nanocarrier systems, for instance for drug delivery or adaptive nanocatalysis, is the partition \chgA{ratio}, which quantifies the uptake of a molecular substance by the polymer matrix. By employing all-atom molecular dynamics simulations,  we study the solvation and partitioning (with respect to bulk water) of small subnanometer-sized solutes in a dense matrix of collapsed Poly(N-isopropylacrylamide) (PNIPAM) polymers above the lower critical solution temperature (LCST) in aqueous solution.  We examine the roles of the solute's polarity and its size on the solubility properties in the thermoresponsive polymer. We show that the transfer free energies of nonpolar solutes from bulk water into the polymer are favorable and scale in a good approximation with the solute's surface area. Even for small solute size variation, partitioning can vary over orders of magnitude.  A polar nature of the solute, on the other hand, generally opposes the transfer, at least for alkyl solutes. Finally, we find a strong correlation between the transfer free energies in the gel and the adsorption free energies on a single extended polymer chain, 
which enables us to relate the partition \chgA{ratios} in the swollen and collapsed state of a PNIPAM gel.
\end{abstract}	

\maketitle
\setlength\arraycolsep{2pt}

\section{Introduction}
Stimuli-responsive hydrogels are one of the most promising types of polymers in the development of new soft functional materials~\cite{klouda2008thermoresponsive}.
Their attractiveness stems mainly from their responsive nature, tunable water content, and rubbery character, as they resemble the features of biological tissue~\cite{peppas1997hydrogels}.
These properties made them a key ingredient in many applications in material science, especially as nanocarrier systems, spanning from drug delivery~\cite{peppas1997hydrogels, stuartNatMat2010, kabanovAngew2009, oh2008development,yingApplicationsSM2011}, catalysis~\cite{jandtCatal2010, ballauffRoyal2009, dzubiellaAngew2012}, biosensing~\cite{yingApplicationsSM2011,ravaineGlucose2006}, thin-film techniques~\cite{yingApplicationsSM2011}, as well as in environmental science~\cite{serpeACS2011}, \chgA{including nanofiltration and water purification}~\cite{nykanen2007phase, chen2014thermo, hyk2018water}.
Undoubtedly, the most studied thermoresponsive polymer is poly(N-isopropylacrylamide) (PNIPAM), which exhibits the volume transition in water close to the human body temperature~\cite{pelton2000, hudsonProgPolySci2004, halperin2015poly}. As a versatile model component, it has served as a prototype for many developments of soft responsive materials~\cite{stuartNatMat2010, ballauffAngew2006, ballauffSmartPolymer2007}.

Clearly, the complete description of a hydrogel material requires the knowledge of many  material parameters.
Yet, the quantity of central importance is the partition \chgA{ratio} $K$, defined as the ratio of the  concentration of the molecular compound in the hydrogel relative to that in water in thermodynamic equilibrium~\cite{leo1971partition, IUPACgoldbook}.
The partition \chgA{ratio} is used to assess the transfer processes, the uptake of a compound by the hydrogel, the permeability of a compound, and its fate in a hydrogel-based nanocarrier~\cite{chiou2005improved, roa2017catalyzed}.
  Unfortunately, accurate experimental measurements of partition \chgA{ratios} present a serious challenge for many materials~\cite{chiou2005improved}. 

Predictions of the partitioning in collapsed hydrogels are notoriously difficult, owing to the crowded environment with complex polymer--water interactions~\cite{merrill1993partitioning}.
 Partition \chgA{ratios} are many times greater than predicted by size-exclusion theories, which assume simple steric hindrance of solutes by the network.
In fact, because of a lower water content, a collapsed hydrogel exhibits a more hydrophobic environment than in the swollen state, therefore the partition \chgA{ratio} of many hydrophobic compounds (typically drugs) can exceed unity (that is, bulk concentration) by orders of magnitude~\cite{palasis1992permeability, guilherme2003hydrogels, molinaPolymer2012}.

Thus, estimations of partition \chgA{ratios} rely, especially in pharmaceutical research,  on various heuristic approaches that are based on statistical analysis of big experimental data collections~\cite{katritzky2000structurally, duffy2000prediction, jorgensen2002prediction}. They typically involve different descriptors, such as the size of the molecule,  hydrogen-bond counts, the Coulomb energy, etc.~\cite{duffy2000prediction, jorgensen2002prediction}
   However, the large numbers of descriptors that are featured in heuristic models and neural networks obscure the physical basis of solvation~\cite{duffy2000prediction}.
Alternatively, for a basic understanding of the solvation and partitioning processes, one needs to resort to 
various computer modeling techniques. They typically comprise multiscale approaches, spanning from coarse-grained implicit-solvent models~\cite{aydt2000swelling, erbas2015energy,kosovan2015modeling, kim2017cosolute, perez2018maximizing, quesada2018direct} to all-atom models~\cite{rodriguez2014direct, vdVegtPCCP2015, perez2015anions, kanduc2017selective}.
Our broader research goal lies in a qualitative molecular-level understanding of thermodynamic and transport properties in PNIPAM-based hydrogels. Employing atomistic, explicit-water simulations, we studied solute partitioning in swollen gels~\cite{kanduc2017selective} and diffusion inside collapsed gels~\cite{kanduc2018diffusion}.

In this study, we adopt the previous models to investigate the solvation and partitioning in collapsed PNIPAM polymers of several polar and nonpolar molecules, featuring small alkanes, simple alcohols, and aromatic molecules.



     
\section{Methods}
\subsection*{Atomistic model}

We set up an atomistic model of collapsed PNIPAM polymer chains in explicit water. The collapsed polymer can also serve as an approximate model for a collapsed hydrogel with very low (few percents) cross-linker concentrations.  We utilize the same simulation model as in our recent work~\cite{kanduc2018diffusion}, which primarily focuses on molecular diffusion.
To briefly recap, the PNIPAM chains are composed of 20 monomeric units with atactic stereochemisty (\ie, with random distribution of monomeric enantiomers along the chain). 
For PNIPAM polymers we adopt the recent OPLS-based force field by Palivec et al.~\cite{palivecheyda2018} with an {\it ad hoc} parametrization of partial charges, which reproduces the thermoresponsive properties much better than the standard OPLS-AA force field, even though the latter one used to be very popular for PNIPAM simulations~\cite{walter-PNIPAM2010, vegtJPCB2011, stevens_Macro2012, mukherji2013coil,  chiessi2016influence, rodriguez2014direct, adroher2017conformation}.
For water we use the SPC/E water model~\cite{spce}, and the OPLS-AA force field~\cite{opls1988, priceOPLS2001} for solute molecules. This force field combination keeps our model on a generic level and captures the hydration properties of solutes in water very well~\cite{hess2006hydration}, as we also demonstrate in \SItext.

\begin{figure*}[t]\begin{center}
\begin{minipage}[b]{0.3\textwidth}\begin{center}
\includegraphics[width=\textwidth]{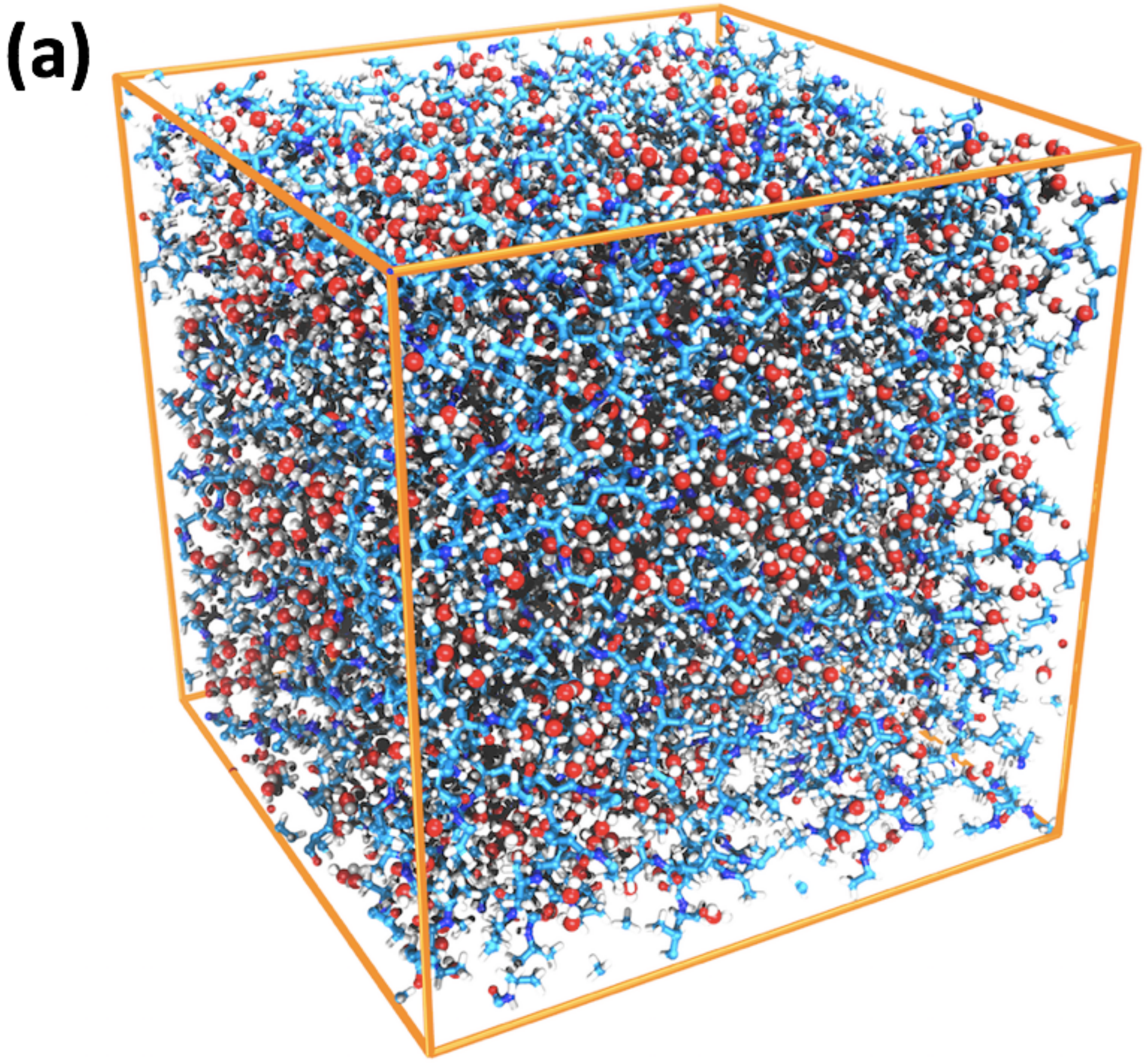}
\end{center}\end{minipage}\hspace{3ex}
\begin{minipage}[b]{0.3\textwidth}\begin{center}
\includegraphics[width=\textwidth]{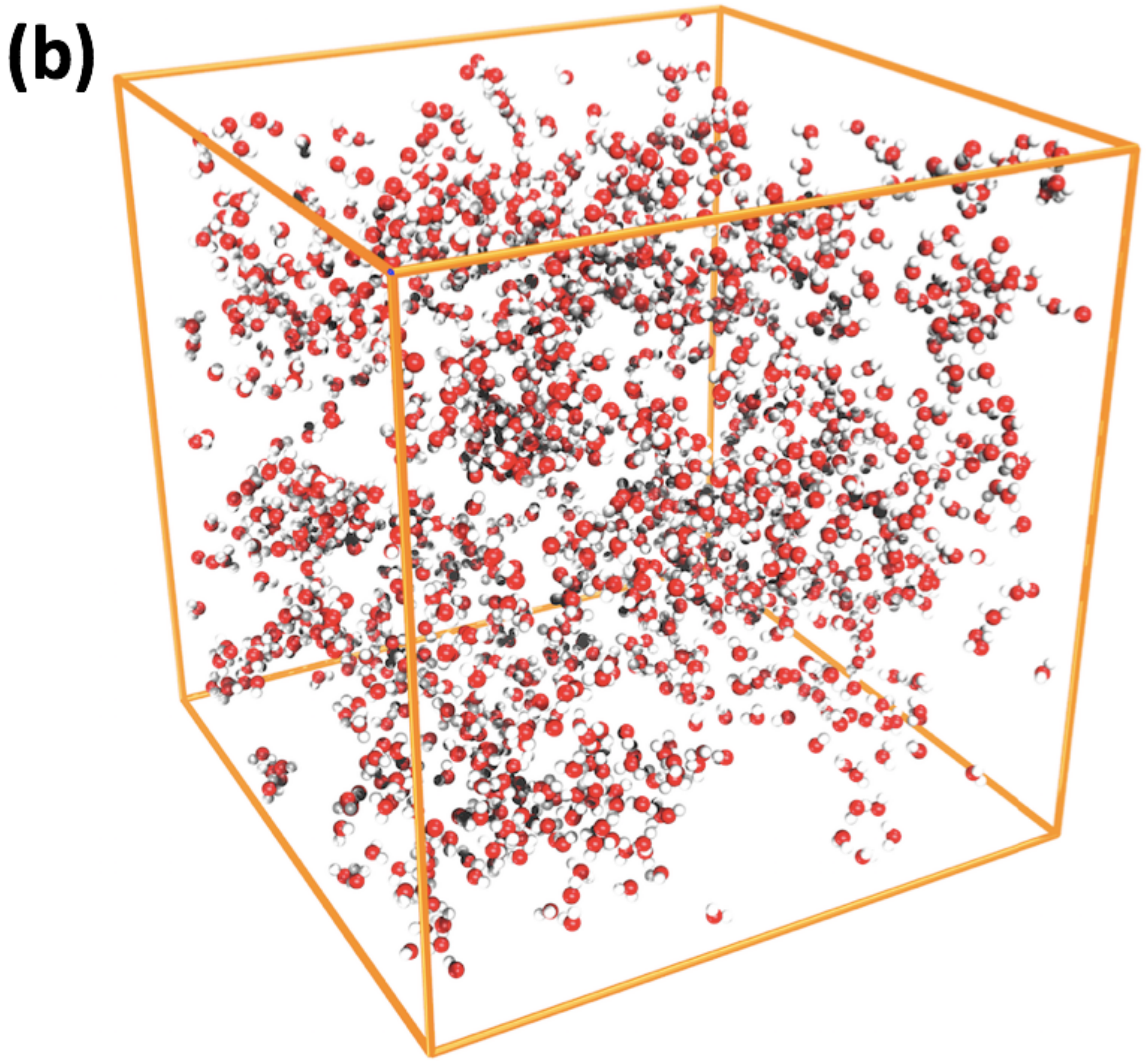}
\end{center}\end{minipage}\hspace{4ex}
\begin{minipage}[b]{0.29\textwidth}\begin{center}
\includegraphics[width=\textwidth]{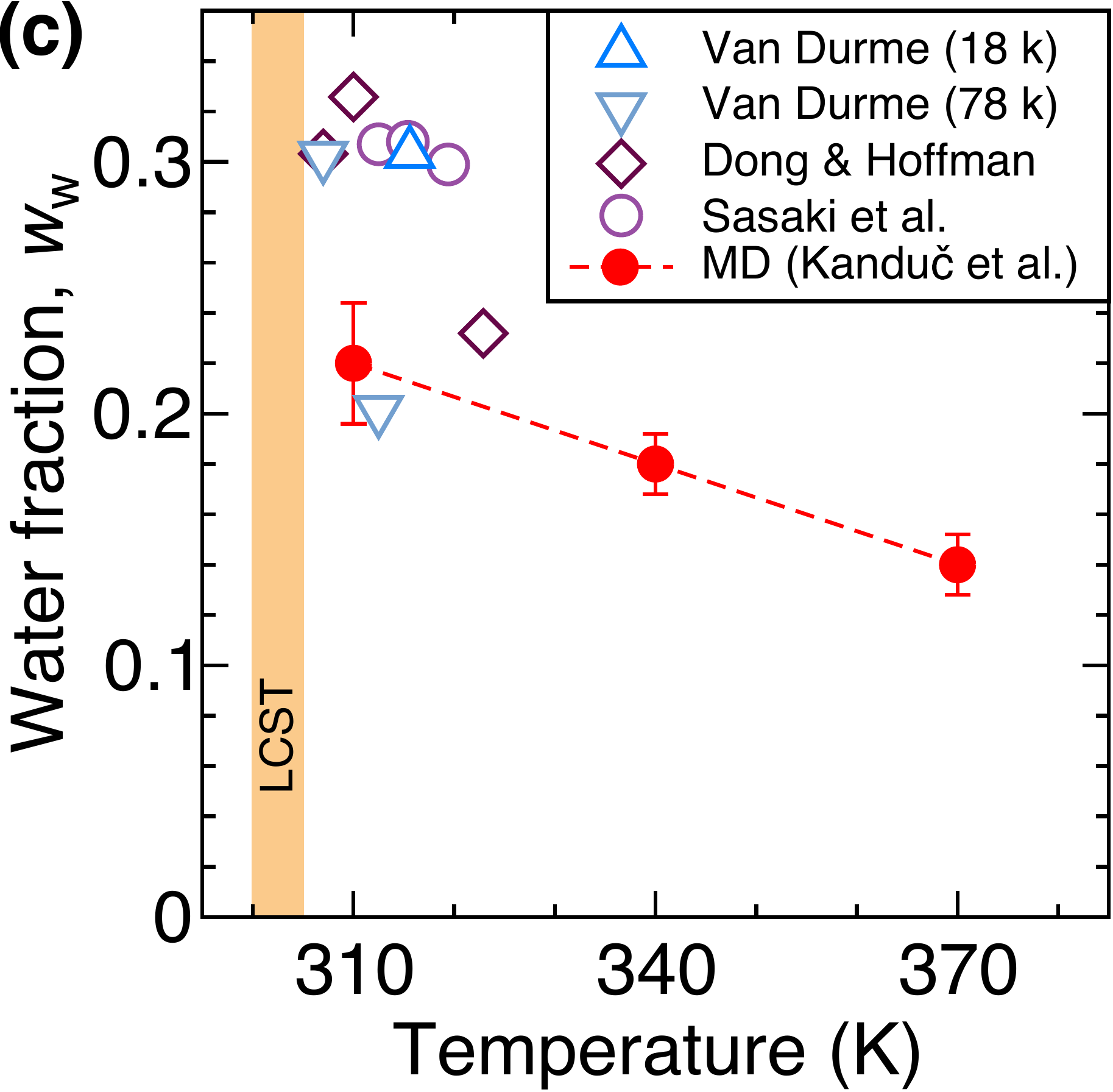}
\end{center}\end{minipage}
\caption{Snapshots of collapsed PNIPAM at 340 K showing (a)~all atoms and (b)~only water molecules.
 (c)~Mass fraction of water in the collapsed PNIPAM that corresponds to chemical equilibrium with bulk water as determined in our previous study~\cite{kanduc2018diffusion} (red circles).
Open symbols show different experimental measurements: Triangles are the PNIPAM/water coexistence (binodal) curves by Van Durme et al.~\cite{vanDurme2004kinetics} for molecular weights of 18 and 78 kDa. Diamonds and circles represent the water amount in PNIPAM hydrogels (\ie, cross-linked networks) by Dong and Hoffman~\cite{dong1990synthesis} and Sasaki et al.~\cite{sasaki1999dielectric}, respectively.}
\label{fig:snap}
\end{center}\end{figure*}

Equilibrated molecular structures of the collapsed PNIPAM polymers with water were taken from the previous study~\cite{kanduc2018diffusion}. First, 48 polymer chains were solvated in excess water above the transition temperature ($\approx$305~K), which caused the polymers to phase separate from the water phase. The amount of water in the polymer precipitate was then used as an input parameter to construct a simulation setup that mimics a bulk of collapsed PNIPAM (\Fig~\ref{fig:snap}a) with employed periodic boundary conditions in all three directions.
The amount of water (\Fig~\ref{fig:snap}b) in the PNIPAM thus corresponds to the equilibrium amount with an external bulk water reservoir, which depends on temperature, as shown in \Fig~\ref{fig:snap}c.

\subsection*{Simulation details}
The molecular dynamics~(MD) simulations were carried out with the GROMACS~5.1 simulation package~\cite{gromacs,gromacs2013} in the constant-pressure (NPT) ensemble, where the box sizes are independently adjusted in order to maintain the external pressure of 1~bar via the Berendsen barostat~\cite{berendsenT} with the time constant of 1~ps. 
The system temperature was maintained by the velocity-rescaling thermostat~\cite{v-rescale} with the time constant of 0.1\,ps.
The Lennard-Jones~(LJ) interactions were truncated at 1.0~nm. Electrostatics was treated using Particle-Mesh-Ewald (PME) methods~\cite{PME1,PME2} with the real-space cutoff of 1.0~nm. 

\subsection{Solvation free energies}
\label{sec:MethodsSolvation}
\begin{figure}\begin{center}
\begin{minipage}[b]{0.48\textwidth}\begin{center}
\includegraphics[width=\textwidth]{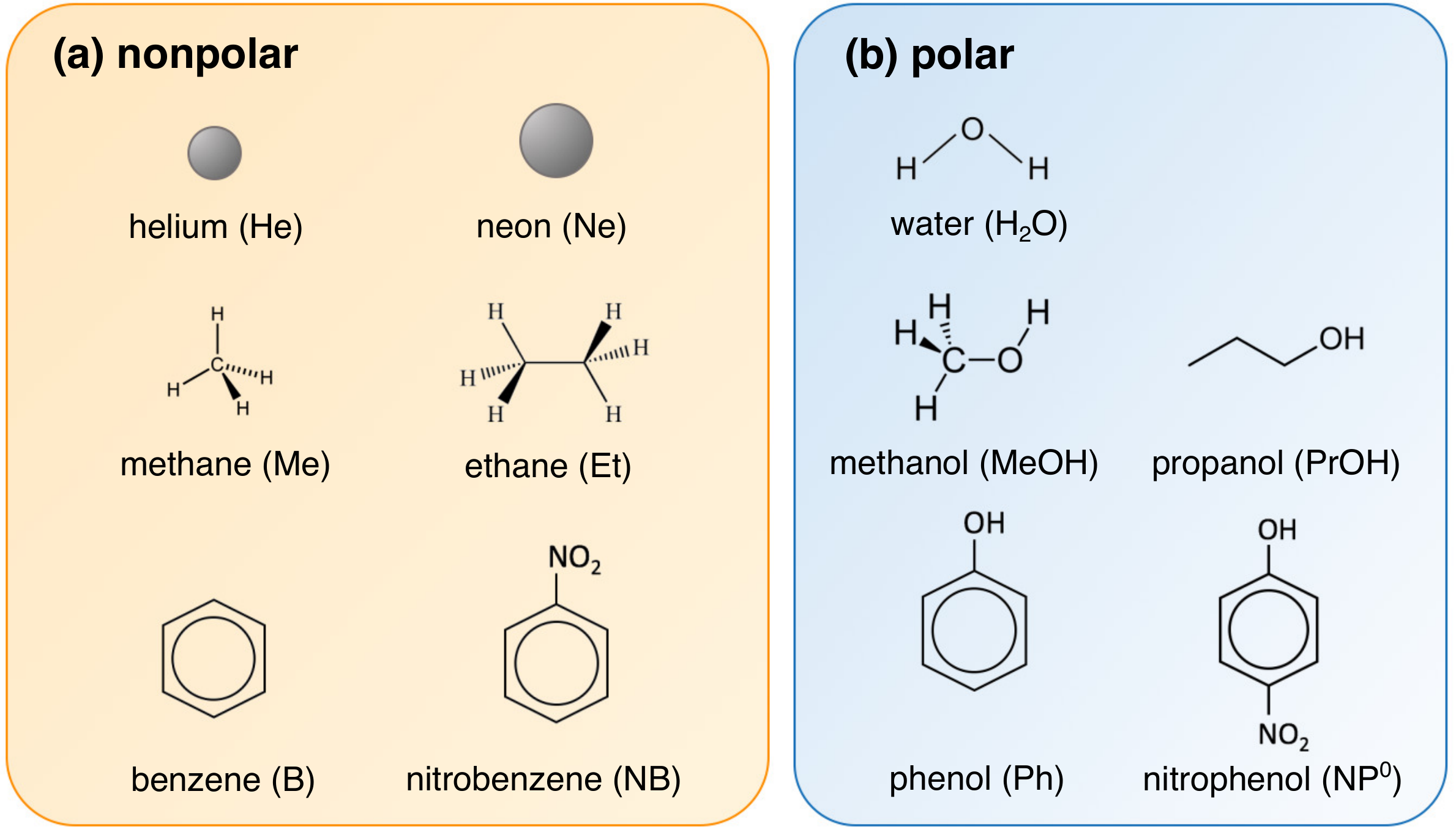}
\end{center}\end{minipage}
\caption{Solute molecules in our study: (a) nonpolar molecules and (b)~polar molecules (characterized by the hydroxyl group).}
\label{fig:solutes}
\end{center}\end{figure}

In order to perform free energy calculations of the solutes (\Fig~\ref{fig:solutes}), we first insert 10--15 solute molecules of the same kind at random positions in an equilibrated system of PNIPAM polymers and water. The system (with fully interacting solutes) then undergoes further equilibration, where the necessary equilibration times of the solutes are estimated based on the crossover times of the solutes to reach the normal diffusion~\cite{kanduc2018diffusion} (such that a solute performs at least a few hopping events and reaches its equilibrium hydration). The equilibration times thus span from several 100~ps for the smallest solutes (He and Ne) and up to several 100~ns for the largest (NP$^0$).

The solvation free energies of the solutes are then computed using the Thermodynamic Integration~(TI) technique~\cite{frenkel1984new}, with the strength of the interaction potential of the solutes as the thermodynamic path. Within this procedure, the solute's partial charges and all LJ interactions between the solute molecule and other molecules are gradually switched off. Introducing a coupling parameter $\lambda\in[0,1]$ that continuously switches the interactions in the Hamiltonian $U(\lambda)$ between the original solute interactions (for $\lambda$\,$=$\,1) and a non-interacting particle (for $\lambda$\,$=$\,0), the solvation free energy is computed as~\cite{frenkel1984new}
\begin{equation}
G^\trm{TI}=\int_0^1\left\langle\frac{\partial U(\lambda)}{\partial\lambda}\right\rangle_\lambda \rmd\lambda
\label{eq:TI}
\end{equation}
The integration is performed in two stages: We first linearly scale down the partial charges of the solute particle, while keeping the LJ interactions intact. In the second stage, we scale the LJ interactions using the ``soft-core'' LJ functions as implemented in GROMACS~\cite{gromacs,gromacs2013} in order to circumvent singularities when the potentials are about to vanish ($\lambda\to0$)~\cite{beutler1994avoiding}. 
We separate the entire TI procedure into 24 individual simulations with equidistant $\lambda$ values for the Coulomb part and likewise 24 simulations for the LJ part.
The simulation time of each individual simulation with given $\lambda$ is 4~ns where the first 0.1--3~ns are discarded from sampling to allow equilibration of the solutes. The estimation of the equilibration time in each individual simulation is based on the drift of the $(\partial U/\partial\lambda)$ output.
All the TI calculations are performed using 2--5 independently equilibrated systems. The final results are averaged over all the particles in a simulation box and over all the systems.
\chgA{
Even though each particle samples only a local phase space during a short TI time window, such an averaging assures adequately sampled values of the free energy.
We have also verified that solute--solute interactions during TI have negligible effects on the free energy evaluation (see \SItext).
}

The solvation free energies for transferring a solute from the vacuum (or the gas phase) into the medium (\ie, bulk water or PNIPAM; $i=\rm w,p$, respectively) are obtained by subtracting the free energy value (\Eq~\ref{eq:TI}) obtained from TI in vacuum from the value in the medium, $G_i=G^\trm{TI}_\trm{(in medium $i$)}- G^\trm{TI}_\trm{(in vacuum)}$. Note that $G^\trm{TI}_\trm{(in vacuum)}$ may be nonzero due to intra-molecular Coulombic and LJ contributions.

\subsection{Molecular surface area}
A convenient and simple way to characterize the effective size of a solute is based on its surface area. 
For our purposes, we consider a solute as a union of fused van der Waals spheres. The {\sl molecular surface area}, $A_\trm m$, is then the envelope area of the fused union of the spheres~\cite{karelson2000molecular}.
This definition is analogous to the concept of the solvent accessible surface area (SASA)~\cite{shrake1973environment, eisenhaber1995double} with vanishing probe radius, $r_0=0$, and in a sense corresponds to the surface area of a `bare' solute. 

\section{Results and Discussion}
\subsection{Water structure in the collapsed polymer}
For convenience of the reader we first briefly recall some basic characteristics of the collapsed PNIPAM polymer with sorbed water, as in detail presented in our previous study~\cite{kanduc2018diffusion}.
We performed simulations at 310, 340, and 370~K (range where PNIPAM is collapsed). Most of our analysis and discussion will be centered around the results at the midpoint temperature of 340~K.

  A typical simulation snapshot at 340~K is shown in \Fig~\ref{fig:snap}a with PNIPAM chains in blue and water molecules in red--white. For better illustration, the water component of this structure is displayed alone in \Fig~\ref{fig:snap}b. Water molecules are far from being uniformly distributed in the polymer phase, but rather structure into nanosized `lacy-like' clusters and pockets.
Additionally, the water clusters do not have a characteristic size but are extremely polydisperse, with a size distribution that roughly follows the power law $P(N_\trm{w})\propto N_\trm{w}^{-1.74}$, where $N_\trm w$ is the number of water molecules in the cluster. The radius of gyration $R_\trm g$ of the clusters scales with the size $N_\trm w$ as $R_\trm g\sim {N_\trm w}^{1/2}$, which is also a characteristic of the random walk~\cite{kanduc2018diffusion}.

The amount of water varies with temperature such that it always matches the chemical equilibrium with a hypothetical external water reservoir. The corresponding amount has been determined in the previous study~\cite{kanduc2018diffusion}, and is shown in \Fig~\ref{fig:snap}c in terms of  mass water fraction, $w_\trm w$, at different temperatures. The mass fraction of water is the ratio between the water mass density inside PNIPAM, $\rho_\trm w^\trm{(in)}$, and the density of the entire PNIPAM phase,  $w_\trm{w}={\rho_\trm w^\trm{(in)}}/\rho_\trm{tot}^\trm{(in)}$.
 The water fraction linearly decreases with rising temperature, which reflects the increasing hydrophobic character of PNIPAM upon heating.
Analogously to the classical partition \chgA{ratio}, we can define the water partition \chgA{ratio} $K_\trm w$ as the density ratio of water inside and outside the polymer.
Since the density of the PNIPAM  phase is roughly the same as the density of bulk water, $\rho_\trm{tot}^\trm{(in)}\approx \rho_\trm w^0$~\cite{kanduc2018diffusion}, it follows that the water partition \chgA{ratio} approximately equals the water mass fraction,
\begin{equation}
K_\trm w=\frac{\rho_\trm{w}^\trm{(in)}}{\rho_\trm{w}^{0}}\approx w_\trm w
\label{eq:Kw}
\end{equation}

\subsection{Distribution of solutes in the PNIPAM phase}
\begin{figure}\begin{center}
\begin{minipage}[b]{0.41\textwidth}\begin{center}
\includegraphics[width=\textwidth]{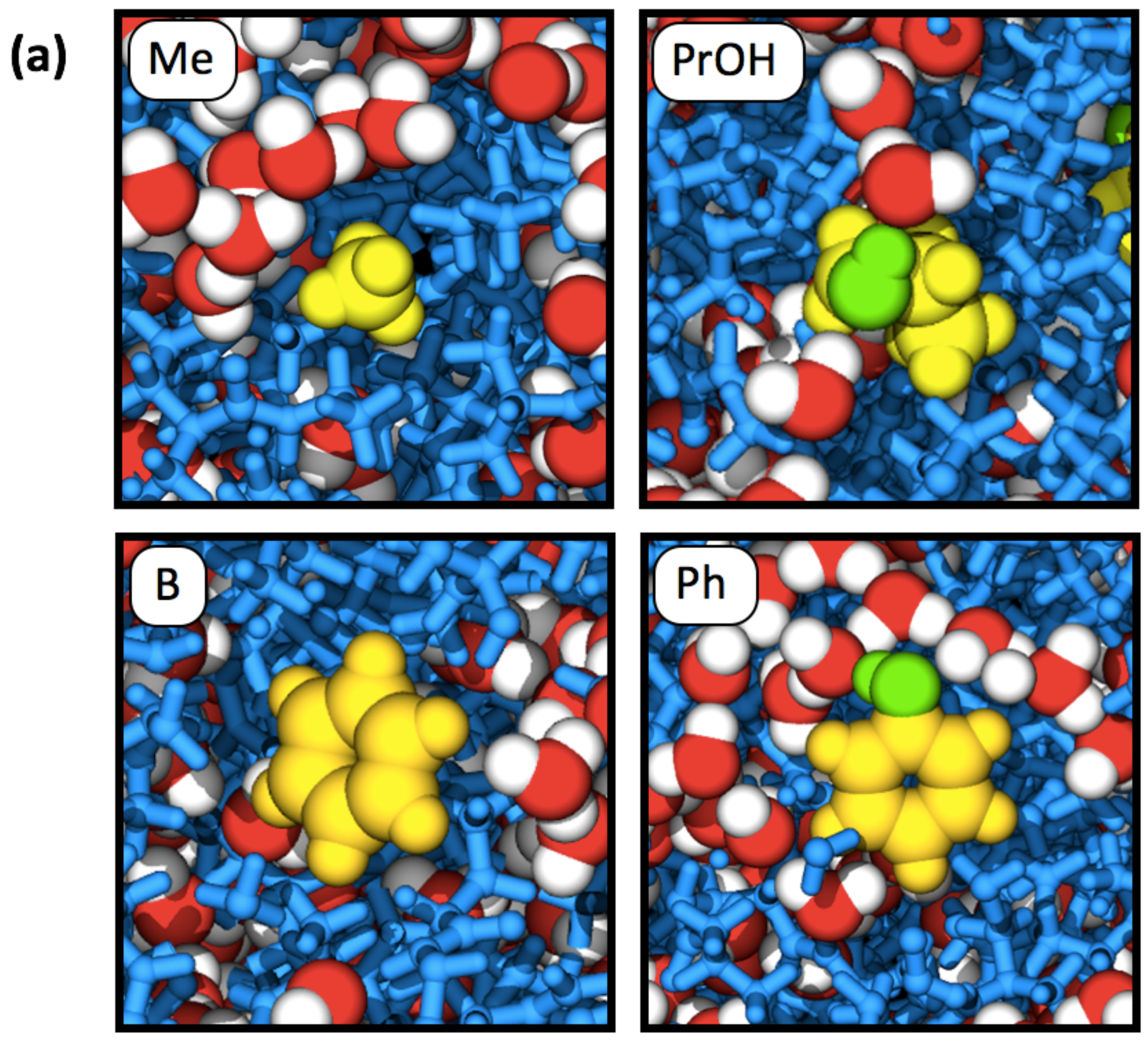}
\end{center}\end{minipage}\vspace{1ex}
\begin{minipage}[b]{0.45\textwidth}\begin{center}
\includegraphics[width=\textwidth]{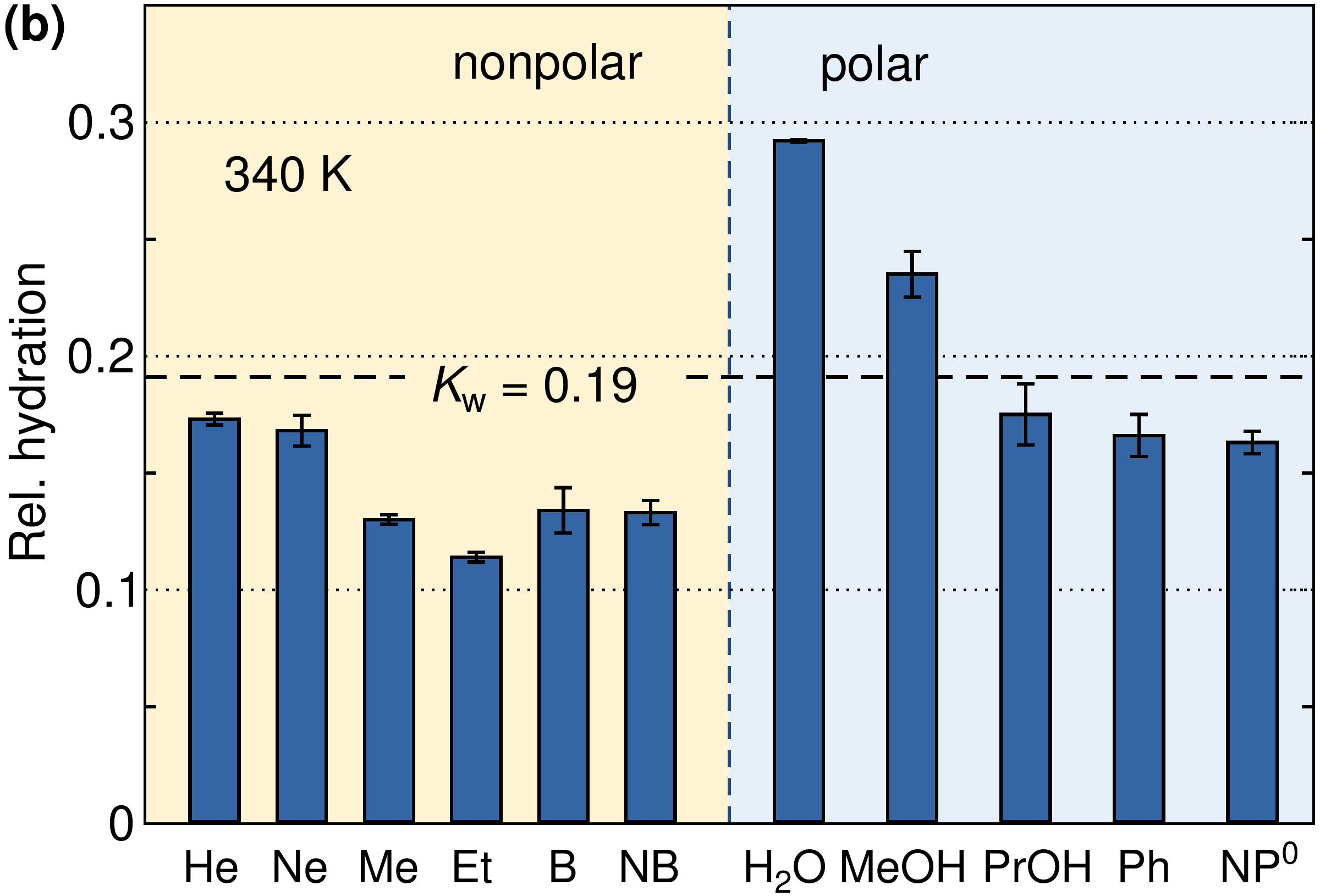}
\end{center}\end{minipage}
\caption{(a) Cross-section snapshots of Me, PrOH, B, and Ph solutes in PNIPAM phase. For clarity are the hydrophobic parts of the solutes shown in yellow, the hydroxyl groups in green, PNIPAM polymers in blue, and water in red--white.
(b)~Relative hydration of solutes in the PNIPAM phase at 340~K with respect to bulk water. The water partition \chgA{ratio} in PNIPAM, $K_\trm w=0.19$ (\Eq~\ref{eq:Kw}) is indicated by a dashed horizontal line and serves as an orientation for the preferred partitioning between water clusters and dry regions of the polymer.}
\label{fig:dehydration}
\end{center}\end{figure}

Due to the aforementioned water--polymer spatial heterogeneity, one can anticipate that solutes are not going to redistribute uniformly throughout the PNIPAM phase, but will be subject to the water--polymer spatial pattern, whereby the nature of the solute (polar vs.\ nonpolar) plays a decisive role. For a qualitative appraisal, we show typical snapshots of several solutes in the PNIPAM phase in \Fig~\ref{fig:dehydration}a. As seen, hydroxyl groups tend to stick to water clusters more willingly than their hydrophobic residues.

In order to examine the nature of the solute distribution quantitatively, we probe the local physical environment into which the solutes settle. For that purpose, we define the hydration number $n_\trm w$ as the number of water molecules residing within a spherical shell of radius $r_\trm{c}=$~0.54~nm (corresponding to the first hydration shell of CH$_4$ in pure water~\cite{kanduc2018diffusion}) around any of the solute's atoms. 
In \Fig~\ref{fig:dehydration}b we show the relative hydration of the solutes, defined as the ratio of the hydration numbers inside PNIPAM and in bulk water, $n_\trm w^\trm{(in)}/n_\trm w^\trm{(out)}$.
A hypothetical solute that would partition uniformly throughout the PNIPAM phase would feature the relative hydration equal to the water partition \chgA{ratio}, $K_\trm w=0.19$ (denoted by a dashed line).
Not surprisingly, nonpolar solutes get significantly more dehydrated when entering the PNIPAM phase, with relative hydration of only 0.11--0.16 relative to water environment. This means, the nonpolar solutes reside preferentially in `dry' regions of the polymer and are expelled from water clusters.
On the contrary, the polar molecules feature on average higher hydration due to the high affinity of the hydroxyl group to water. Clearly, their polar character is diminishing with the size of the nonpolar residue. Thus, polar molecules with larger nonpolar residues (PrOH, Ph, and NP$^0$) behave as being nonpolar in that sense, with the relative hydration below $K_\trm w$.



\subsection{Solvation free energies}

\begin{figure}[t]\begin{center}
\begin{minipage}[b]{0.44\textwidth}\begin{center}
\includegraphics[width=\textwidth]{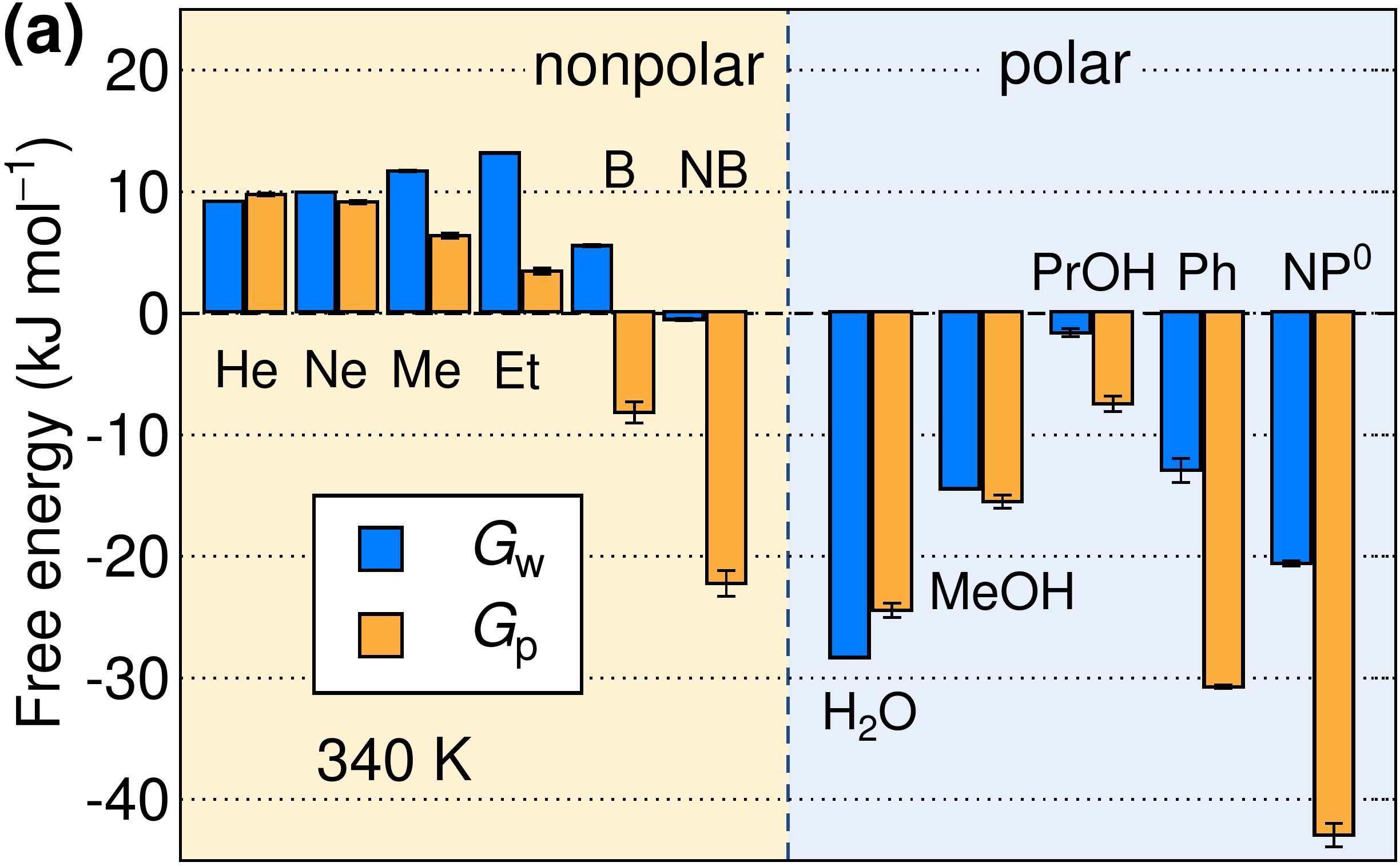}
\end{center}\end{minipage}\vspace{3ex}
\begin{minipage}[b]{0.38\textwidth}\begin{center} 
\includegraphics[width=\textwidth]{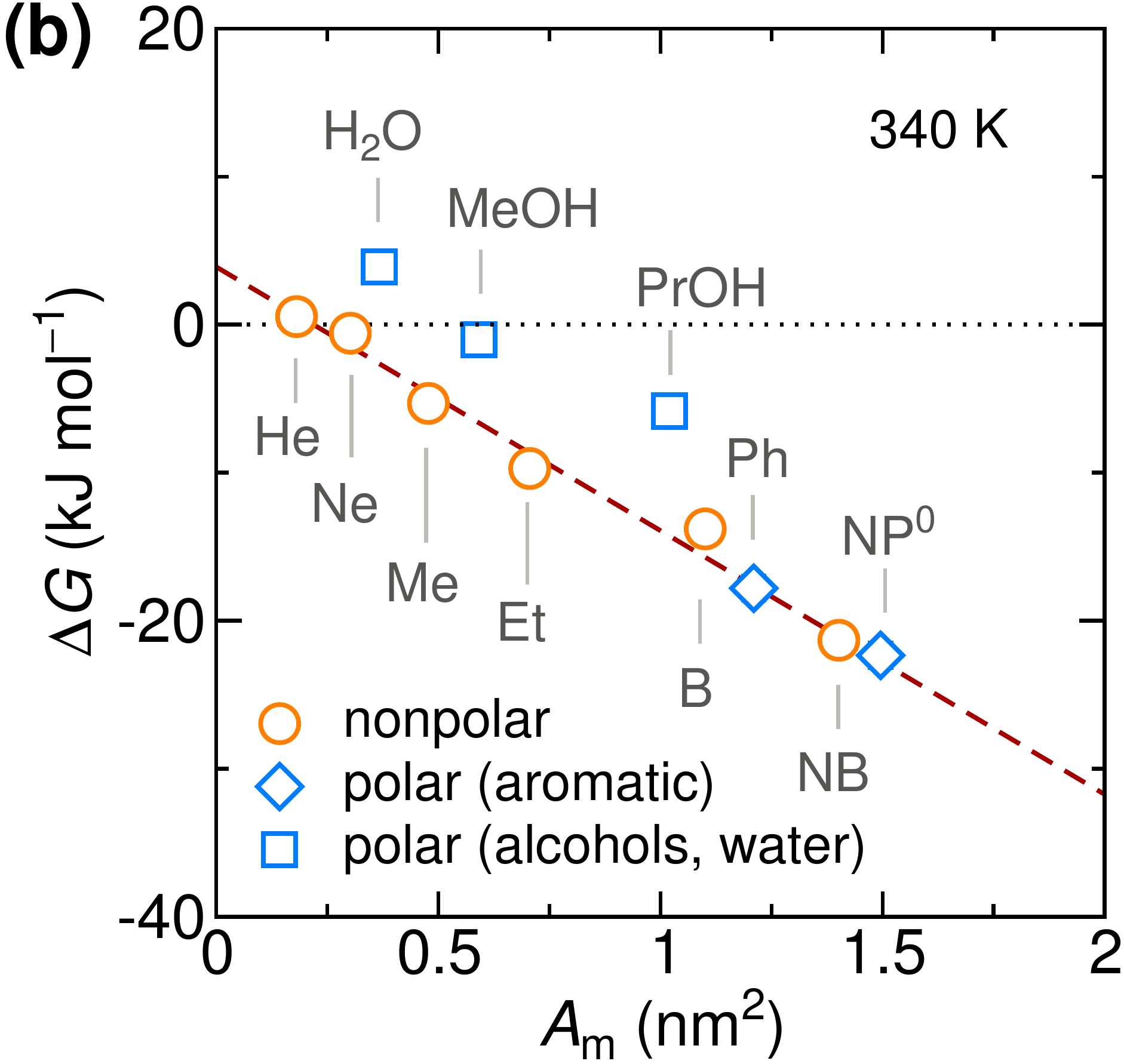}
\end{center}\end{minipage}
\caption{(a) Solvation free energies of solutes in water $G_\trm{w}$ (blue) and in PNIPAM $G_\trm{p}$ (orange) at 340~K.
(b)~Transfer free energies from water to PNIPAM, $\Delta G=G_\trm{p}-G_\trm{w}$, plotted as a function of the molecular surface area.
The dashed line is a fit of \Eq~\ref{eq:Ga2} to the data points of nonpolar solutes.
 }
\label{fig:free}
\end{center}\end{figure}

We now turn to the solvation free energies, as they are central quantities for resolving the partition \chgA{ratios} of the solutes. The solvation free energies for the transfer of a solute from vacuum (or ideal gas phase) into water ($G_\trm{w}$) and collapsed PNIPAM phase ($G_\trm{p}$), are plotted in  \Fig~\ref{fig:free}a.

 Universally, the free energies of nonpolar solutes in water are positive, $G_\trm w>0$, which indicates their low solubility in water. A comparison of the $G_\trm w$ values with experiments shows reasonable agreement (see \SItext), which validates the combination of the used force fields for our needs.
The polar solutes, on the other hand, exhibit significantly negative solvation free energies in water, $G_\trm w<0$, due to the expected formation of hydrogen bonds between the hydroxyl group and water.

The solvation free energies in PNIPAM, $G_\trm p$, are shown in the same plot by orange bars. They are positive and comparable to $G_\trm w$ for small nonpolar molecules, whereas for larger nonpolar as well as for polar molecules, $G_\trm p$ values become negative.
A glance at \Fig~\ref{fig:free}a suggests considerable specificity of the interactions, without a simple rule that can be deduced only by looking at the solutes' molecular structure.
Clearly, the free energies reflect the complexity of detailed chemical structures.

In the remaining part of the paper we turn our attention to the {\sl transfer} free energies, that is, the difference between the solvation free energies in PNIPAM and in water,
\begin{equation}
\Delta G=G_\trm{p}-G_\trm{w}
\end{equation}
 This quantity thus corresponds to the work needed to transfer a solute from bulk water into the PNIPAM phase, and is related to the partition \chgA{ratio in the limit of infinite dilution} as $K=\exp(-{\Delta G/\kB T})$. See also the in-depth discussion on $K$ later on. 
\Fig~\ref{fig:free}b shows $\Delta G$ versus the molecular surface area $A_\trm m$ (see Methods section for the definition of  $A_\trm m$) of the solutes, which reveals a clear linear dependence for the nonpolar solutes. \chgA{A linear scaling is well known for the hydration free energies of large hydrophobes  in bulk water. However, the linearity in our heterogeneous polymer--water medium is surprising.}
The results can be conveniently described in terms of an effective {\sl molecular surface tension}, $\gamma_\trm{m}$~\cite{tanford1979interfacial, ashbaugh2006colloquium},
\begin{equation}
\Delta G=\Delta G_0+\gamma_\trm{m} {A_\trm{m}}
\label{eq:Ga2}
\end{equation}
Note that $\gamma_\trm{m}$ is not equivalent to the free energy of creating a flat macroscopic interface, and can therefore also be negative. 
The fit of \Eq~\ref{eq:Ga2} to the nonpolar solutes (dashed line in \Fig~\ref{fig:free}b) gives the value $\gamma_\trm{m}=-18(1)$~kJ\,mol$^{-1}$\,nm$^{-2}$. 
This value of course depends on the definition of the surface area, but the linear dependence (\Eq~\ref{eq:Ga2}), as it turns out, is a quite robust feature against  the size definition (see \SItext\ for the sensitivity of $\gamma_\trm{m}$ on the definition of the surface area).
Furthermore, the parameter $\Delta G_0$, which also depends on the choice of the solute size, represents an offset in the linear dependence and does hence not contain a significant physical information.

Turning now to the polar molecules, we notice that both alcohols and water exhibit transfer free energies that are by about 7~kJ/mol above the trend of the nonpolar solutes.
This is in line with experimental observations that hydrophobic dyes tend to partition more in PNIPAM-like networks than hydrophilic ones~\cite{guilherme2003hydrogels}.
But, on the other hand, both polar aromatic molecules (Ph and NP$^0$) behave very similarly to the nonpolar solutes instead.
Namely hydroxyl groups in aromatic molecules have slightly different character than in alkyls chains, owing to the mesomeric and inductive effects of the phenyl group~\cite{hansch1991survey}. The phenyl group allows non-bonding electrons on the oxygen to partially delocalize into the ring via $\pi$  bonding, thus decreasing the electron density in the O--H bond.
These small specific effects in the hydroxyl groups thus cause that the polar aromatic solutes behave more alike to the nonpolar ones rather than to polar alkyl molecules (\ie, alcohols).


It is somehow remarkable that also the water molecule (\ie, the solvent), perfectly fits the linear scheme of alcohols. Of course, the transfer free energy of a water molecule, which is $\Delta G=3.9$~kJ/mol, represents the excess free energy for transferring a water molecule from bulk into an already fully hydrated PNIPAM. It should therefore not be confused with the entire free energy of the system, which apart from the excess part contains also the ideal-gas contribution.

\subsection{Partitioning and temperature effects}
\begin{figure}[h]\begin{center}
\begin{minipage}[b]{0.38\textwidth}\begin{center}
\includegraphics[width=\textwidth]{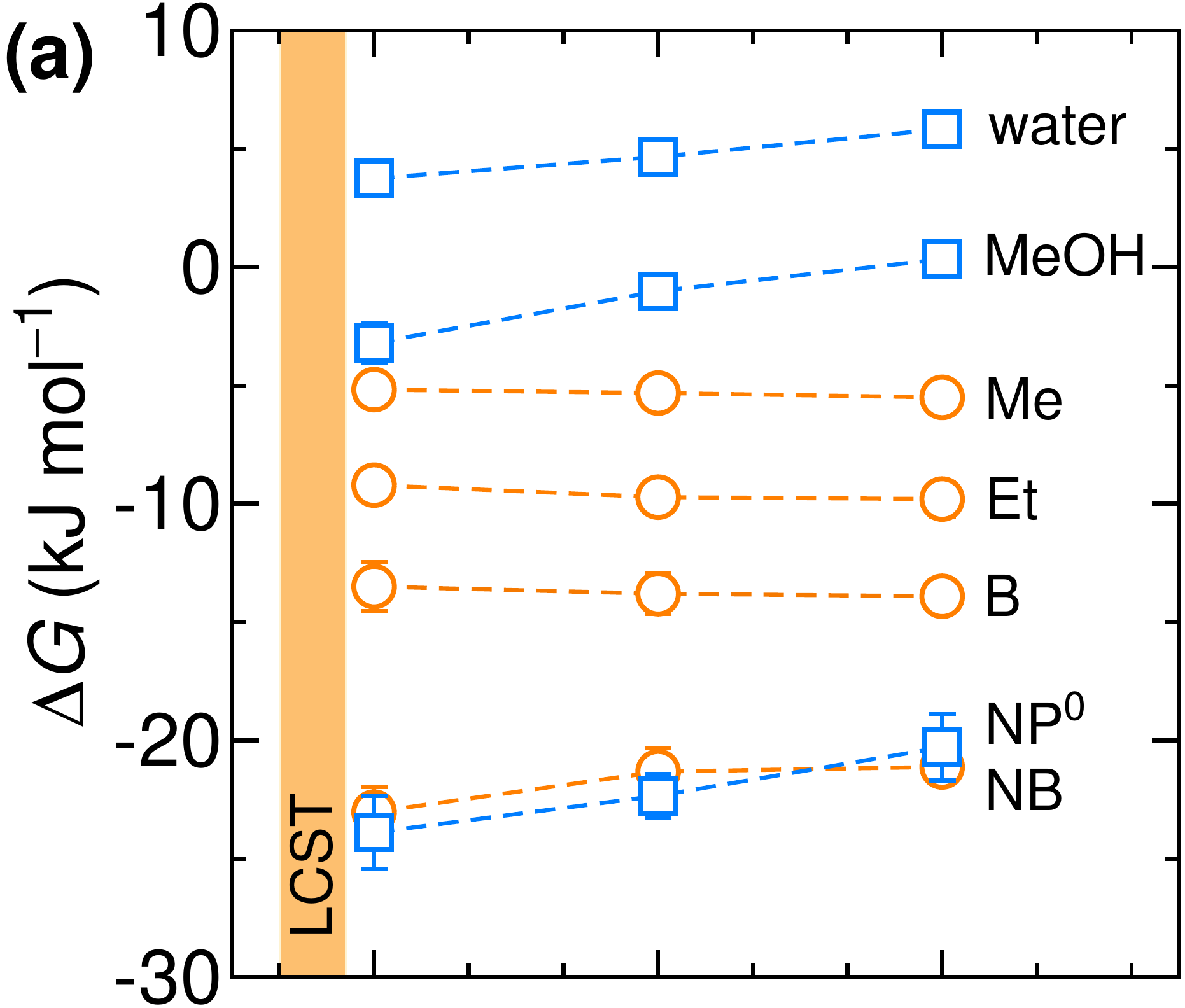}
\end{center}\end{minipage}
\begin{minipage}[b]{0.38\textwidth}\begin{center}
\includegraphics[width=\textwidth]{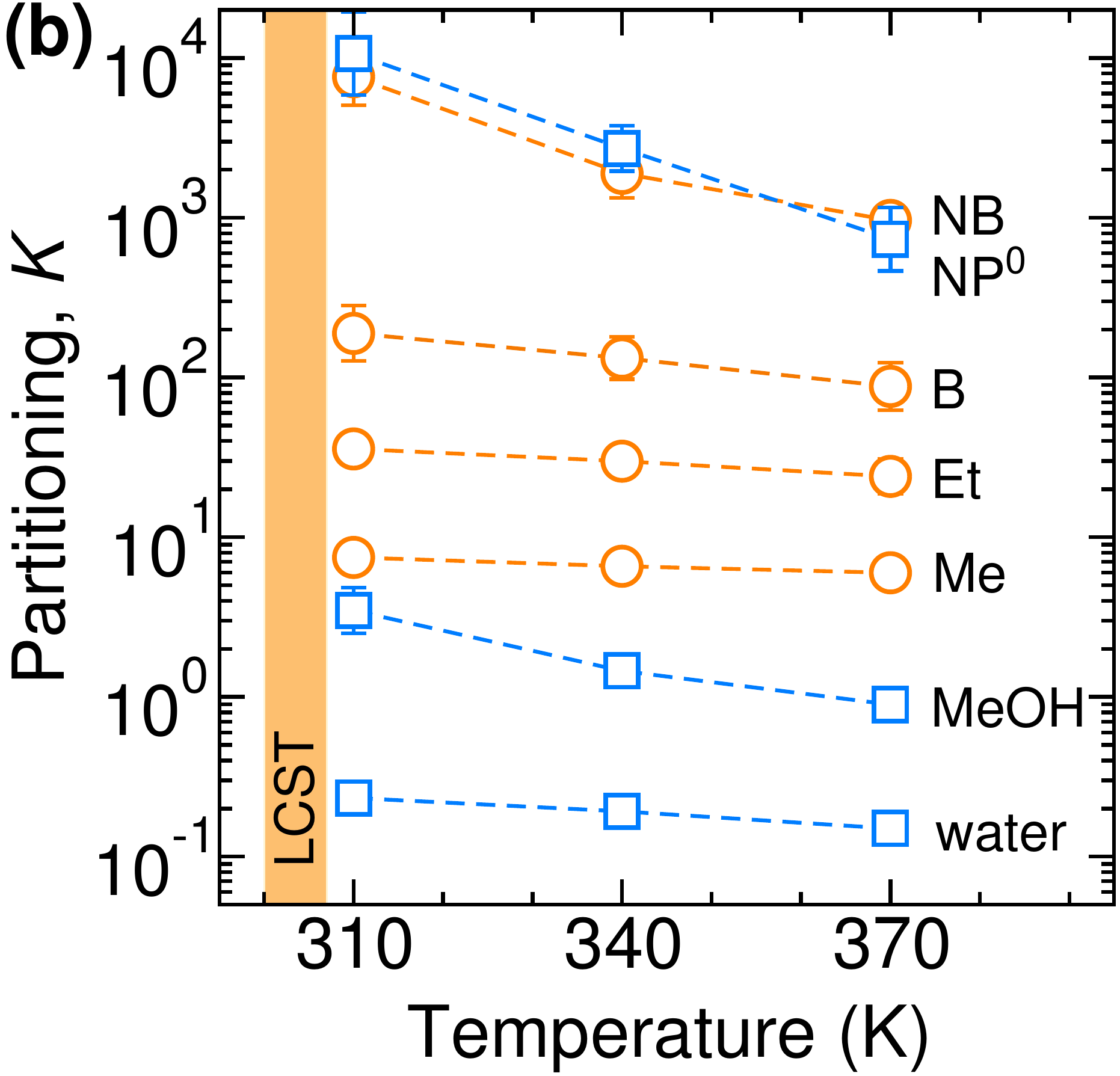}
\end{center}\end{minipage}
\caption{(a) Transfer free energies $\Delta G$ of several solutes as a function of temperature. (b)~The PNIPAM--water partition \chgA{ratio} $K$ (computed via \Eq~\ref{eq:K}) of the solutes from (a) as a function of temperature. Water content in PNIPAM corresponds to the equilibrium value with bulk water reservoir at each temperature (cf. \Fig~\ref{fig:snap}c).
 }
\label{fig:K}
\end{center}\end{figure}

When heating or cooling a collapsed PNIPAM, the water content inside adjusts such that it stays in chemical equilibrium with bulk water (cf.\ \Fig~\ref{fig:snap}c).
Temperature-dependent transfer free energies $\Delta G$ are shown in \Fig~\ref{fig:K}a for several solutes.
Small nonpolar solutes (not NB) show an enhanced solubility (decrease in $\Delta G$) in PNIPAM upon heating.
On the other hand, all polar molecules exhibit a weakening of solvation (increase in $\Delta G$) upon heating. We will return to the explanation of these results later on.

A central quantity of nanocarrier systems in general is the gel--water partition \chgA{ratio} $K$,
 defined as the ratio of the equilibrium concentration in the gel to that in the solution,
 $K=c^\trm{(in)}/c^\trm{(out)}$, and is \chgA{in the limit of infinite dilution} directly related to the transfer free energy $\Delta G$ as
\begin{equation}
K=\exp\left(-\frac{\Delta G}{\kB T}\right)
\label{eq:K}
\end{equation}
Figure~\ref{fig:K}b shows partition \chgA{ratios} that follow from the calculated $\Delta G$ (shown in panel a).
The linear scaling of the transfer free energy with the effective solute surface area (\Eq~\ref{eq:Ga2}) thus results in an almost five-orders-of-magnitude wide span of the partition \chgA{ratios}. Water and methanol feature $K\sim0.1$--$1$, small nonpolar molecules (Me, Et) exhibit $K\sim 10^2$, whereas the partitioning of the largest molecules (NB and NP$^0$) reaches $K\sim 10^4$.

As opposed to mixed temperature dependencies of $\Delta G$ with temperature, $K$ decreases with temperature for all the shown solutes, owing to the additional $1/T$ factor in the exponent of \Eq~\ref{eq:K}.
In fact, the temperature dependence of  $\Delta G$ is related to the transfer entropy $\Delta S$,
\begin{equation}
\left(\frac{\partial \Delta G}{\partial T}\right)_p=-\Delta S
\label{eq:dS}
\end{equation}
On the other hand, the temperature dependence of $K$ is associated with  
the transfer enthalpy $\Delta H$ via the well-known van~'t Hoff equation,
\begin{equation}
\frac{\partial\, \ln K}{\partial T}=\frac{\Delta H}{\kB T^2 }
\label{eq:dK}
\end{equation}
The decrease of $K$ with temperature in our simulations thus indicates exothermic solvation ($\Delta H<0$) in the PNIPAM phase.

\begin{figure*}\begin{center}
\begin{minipage}[b]{0.35\textwidth}\begin{center}
\includegraphics[width=\textwidth]{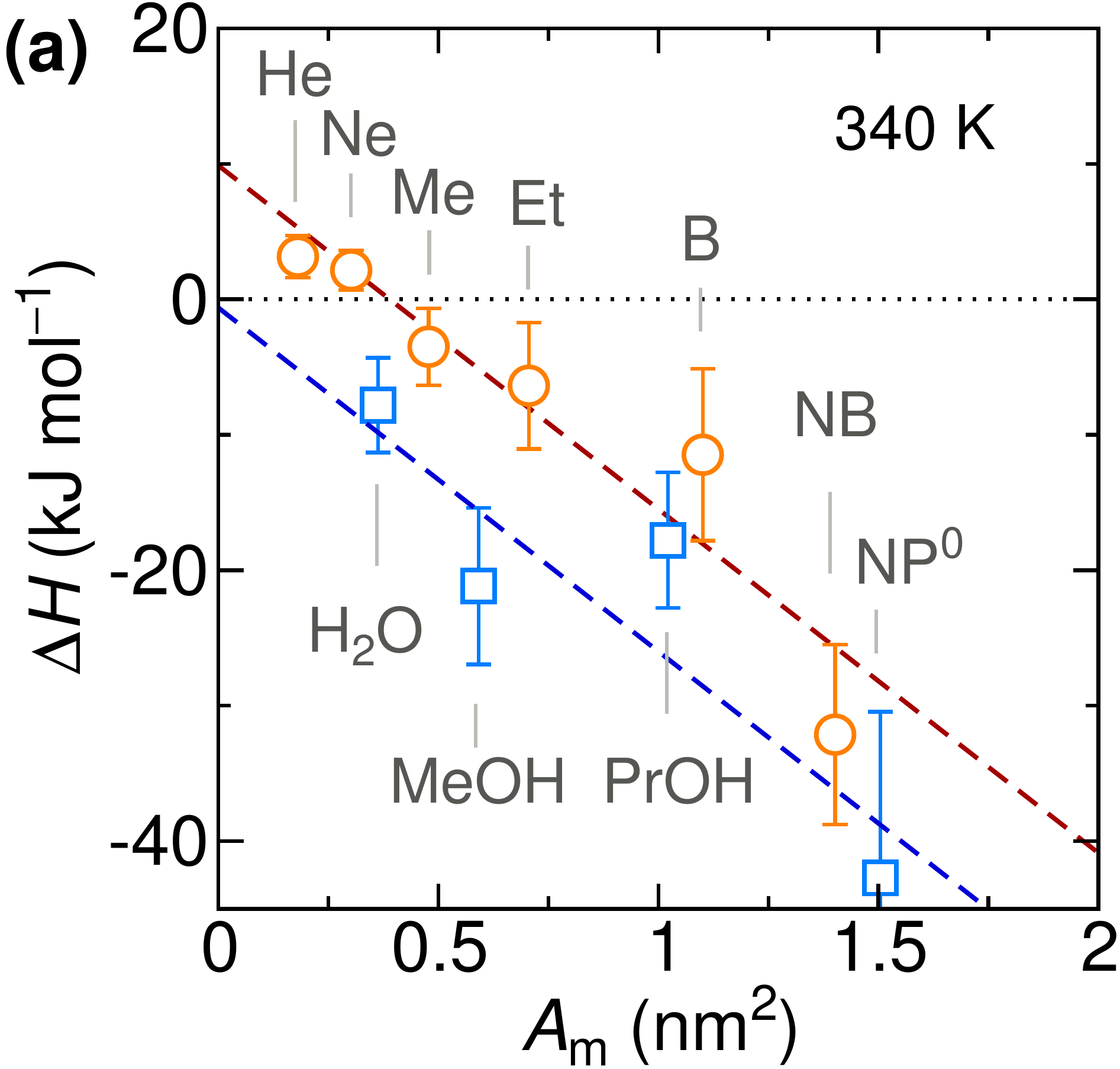}
\end{center}\end{minipage}\hspace{4ex}
\begin{minipage}[b]{0.35\textwidth}\begin{center}
\includegraphics[width=\textwidth]{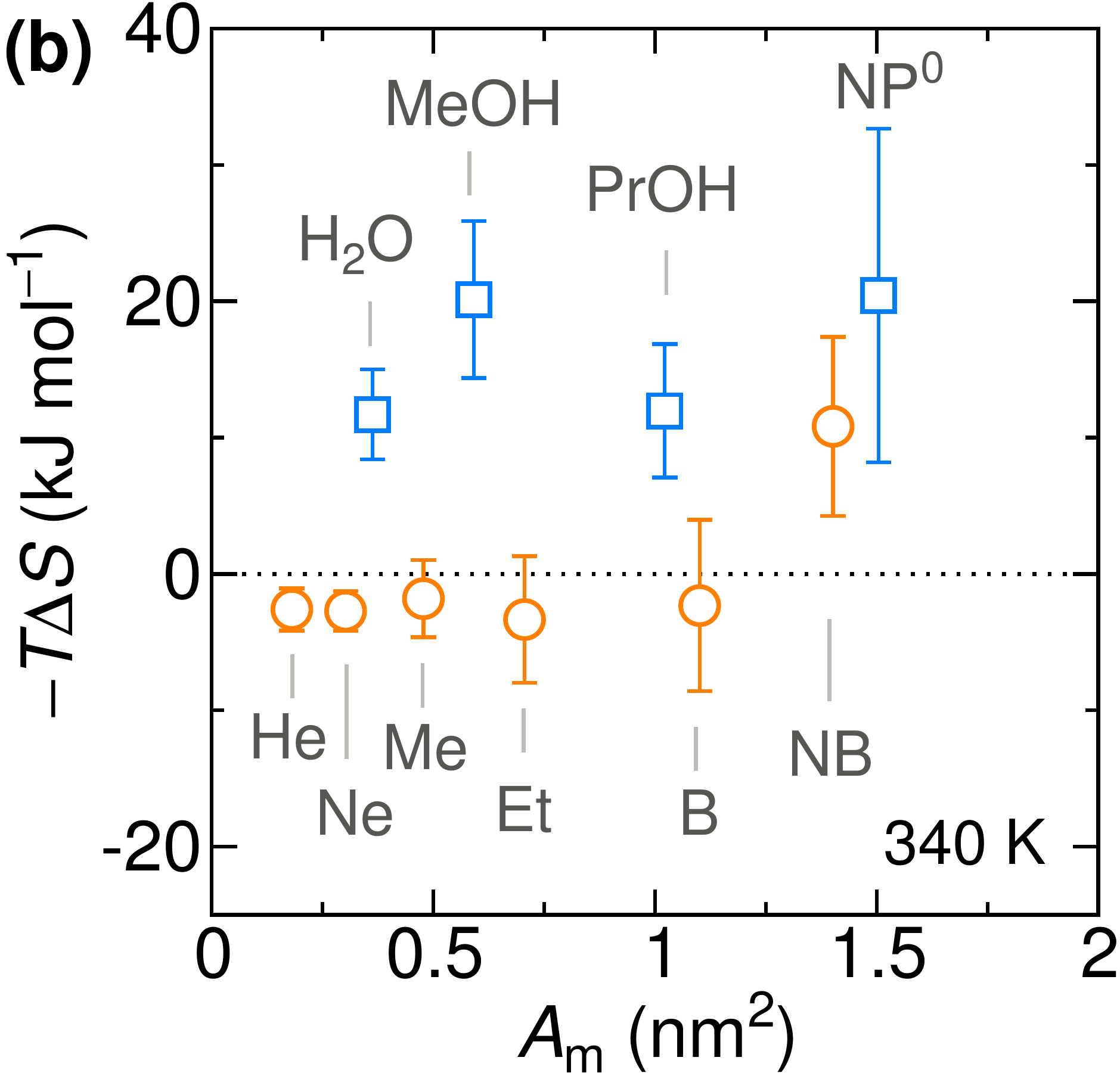} 
\end{center}\end{minipage}
\caption{Decomposition of the transfer free energy $\Delta G$ into (a) the enthalpic and (b) entropic contributions at 340~K.}
\label{fig:decomposition}
\end{center}\end{figure*}

Using \Eqs~\ref{eq:dK} and \ref{eq:dS}, we evaluate the  transfer enthalpies and entropies, respectively, which are plotted in \Fig~\ref{fig:decomposition}.
The enthalpies are negative, except for the smallest solutes (He and Ne), as already implied from the negative trends of $K$ vs.\ $T$ in \Fig~\ref{fig:K}b.
Interestingly, they scale roughly linearly with the particle surface area, and can be fitted to the relation
\begin{equation}
\Delta H=\Delta H_0+h_\trm{m}{A_\trm m} 
\end{equation}
The proportionality factor $h_\trm{m}$ here denotes (in analogy to \Eq~\ref{eq:Ga2}) the {\it surface enthalpy}. A fit to the nonpolar solutes (orange dashed line) yields $h_\trm{m}=-25(5)$~kJ\,mol$^{-1}$\,nm$^{-2}$. 
The polar solutes lie consistently below the nonpolar ones by around $-10$~kJ/mol (as indicated by the blue dashed line with the same $h_\trm{m}$ as for the nonpolar solutes).

The entropic contribution (\Fig~\ref{fig:decomposition}b) for small nonpolar solutes (apart from NB)  is small, slightly negative and almost independent of the size.  This can be partially explained in terms of the unfavorable transfer entropy of small hydrophobes into water~\cite{ashbaugh2006colloquium}. When a solute is transferred from water to more hydrophobic PNIPAM, this results into a weak but favorable free energy component. However, the entropy being so small in magnitude and independent of size is remarkable. We believe also that water molecules with non-saturated hydrogen bonds removed from hydrophobic cavities inside the polymer upon insertion contribute significantly to the free energy, in particular with favorable enthalpy and unfavorable entropy contributions~\cite{setny2010can, weiss2017principles}.
This enthalpy-driven hydrophobic solvation can also be envisioned as 
a complementary association of apolar (concave) pocket--(convex) ligand binding, whose thermodynamic signature hardly depends on the particular pocket/ligand geometry~\cite{dzubiella2011interface}.

For polar solutes, on the other hand, the unfavorable entropy may be related to their restricted (especially rotational) degrees of freedom in the PNIPAM phase. The hydrogen bonds formed between a solute, either with sorbed water or with polymer chains, act as anchors to their motion. In this respect, also a large NB (although classified as nonpolar) can actually form hydrogen bonds via the NO$_2$ group and thus its transfer entropy is in the midway between other nonpolar solutes and the polar ones.


\subsection{Adsorption on an elongated chain}
\begin{figure}[h!]\begin{center}
\begin{minipage}[b]{0.39\textwidth}\begin{center}
\includegraphics[width=\textwidth]{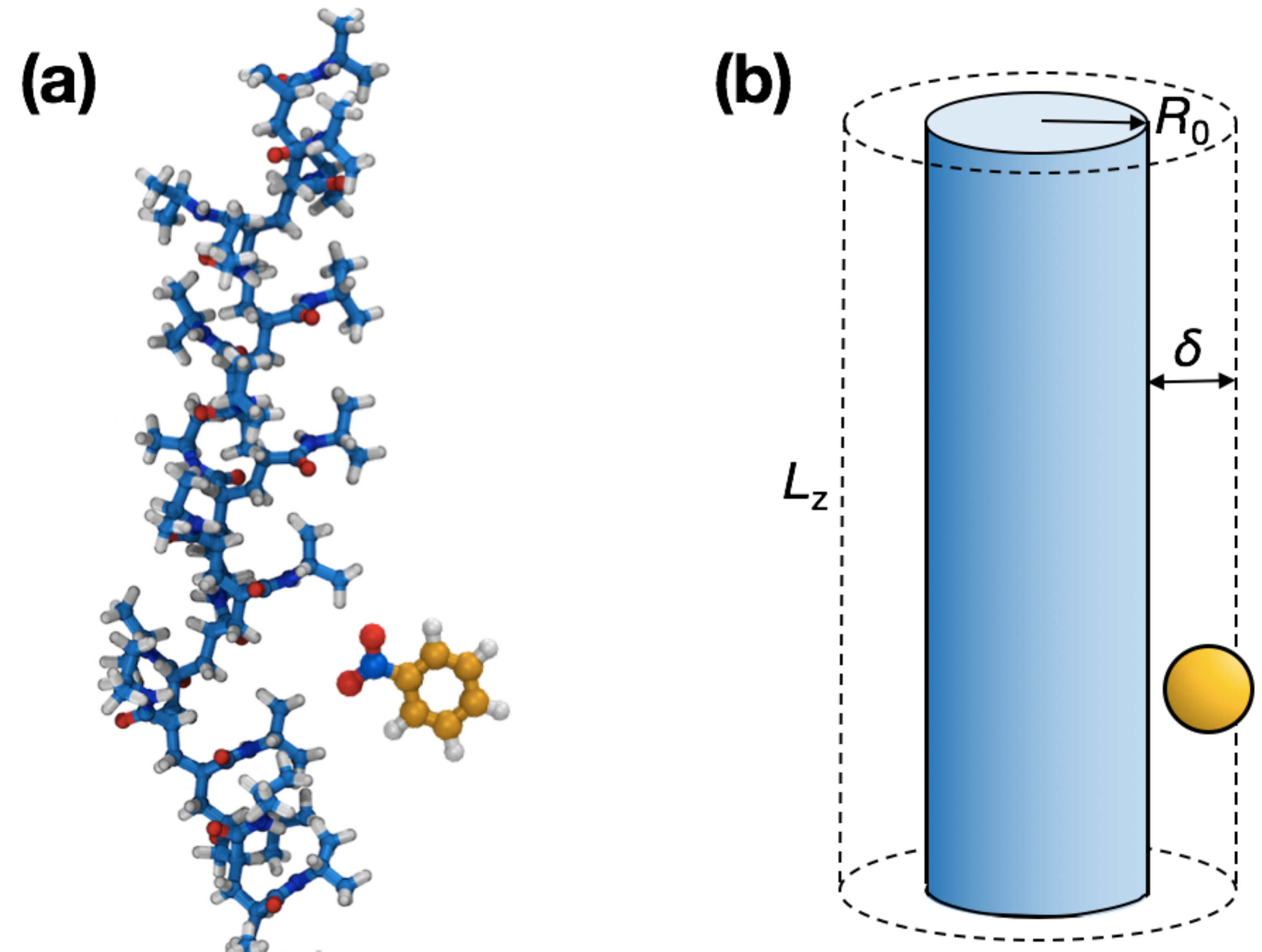}
\end{center}\end{minipage}
\begin{minipage}[b]{0.49\textwidth}\begin{center}
\includegraphics[width=\textwidth]{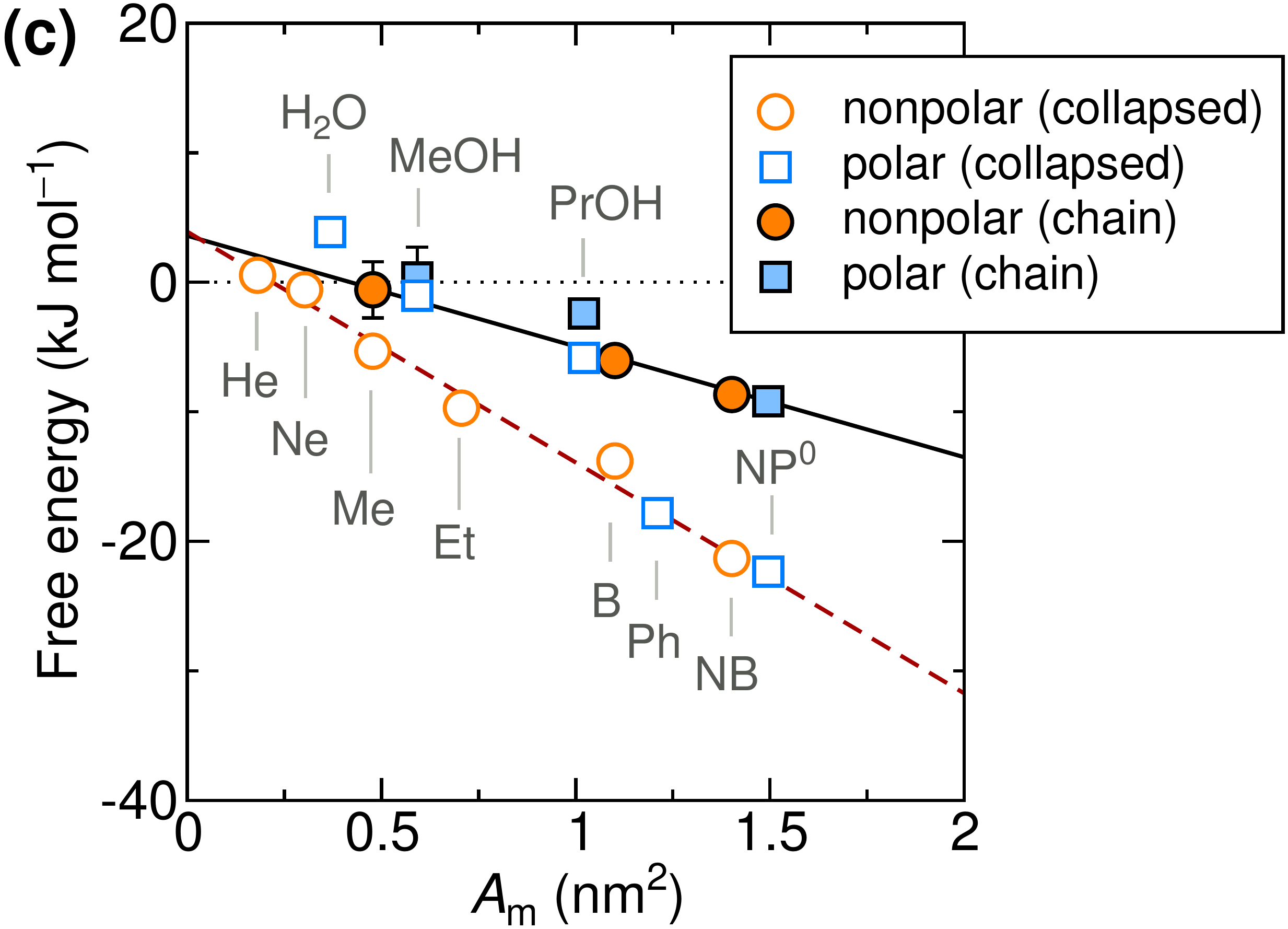}
\end{center}\end{minipage}
\caption{(a) Snapshot of a single elongated PNIPAM chain with a nitrobenzene molecule about to adsorb. (b)~Continuum representation of the chain with radius $R_0$ and an adsorbing region of a thickness $\delta$.
(c)~Comparison of relative solvation free energies in a collapsed PNIPAM (same as in \Fig~\ref{fig:free}b) and the adsorption energies on a single PNIPAM chain evaluated from Ref.~\cite{kanduc2017selective} via \Eq~\ref{eq:Gsingle}.
The lines are fits of \Eq~\ref{eq:Ga2} to the data points of nonpolar solutes.
 }
\label{fig:single}
\end{center}\end{figure}

The atomistic model utilized in this study serves us to obtain an insight into the solvation in a collapsed hydrogel, above the volume transition temperature. Below the transition temperature, a cross-linked PNIPAM hydrogel swells and increases the amount of water typically by an order of magnitude, such that neighboring chains tend to move significantly far apart. A simple model system of a swollen gel was investigated by us in a previous study~\cite{kanduc2017selective}, where we assessed adsorption properties of solutes on a single elongated PNIPAM chain (a snapshot shown in \Fig~\ref{fig:single}a).
It is now interesting to compare solvation in the collapsed and the swollen states of PNIPAM.

The total particle adsorption  $\Delta N$ on a single chain is proportional to the bulk concentration of the solutes, $c^\trm{(out)}$, as $\Delta N=\Gamma' L_z c^\trm{(out)}$, where $L_z$ is the projected chain length and $\Gamma'$ stands for the adsorption coefficient~\cite{kanduc2017selective}. 
\chgA{Note that $\Gamma'$ depends on the chain elongation $\lambda$ (not to be confused with the coupling parameter in TI), defined as the ratio of the projected chain length and the contour length $L_\trm c$ of the chain, $\lambda=L_z/L_\trm c$. }
The adsorption free energy for this system is subject to the definition of bound and unbound states. In a simplified picture (\Fig~\ref{fig:single}b), we can approximate the concentration of the adsorbed solutes in the vicinity of the chain by assuming an effective radius of the chain, $R_0=0.5$~nm (determined as the Gibbs dividing surface of water~\cite{kanduc2017selective}), and specifying an effective adsorption region around the chain cylinder with the width of a typical solute size, $\delta=0.25$~nm.
With that, the concentration of adsorbed solutes expresses as $c_\trm{ads}=\Delta N/(2\pi R_0\delta L_z)$.
The adsorption free energy on a single chain can then be computed as $G_\trm{ads}=-\kB T\,\ln (c_\trm{ads}/c^\trm{(out)})$, which can be written in terms of the adsorption coefficient $\Gamma'$ as
\begin{equation}
G_\trm{ads}=\kB T\, \ln(2\pi R_0\delta)-\kB T\, \ln\, \Gamma'
\label{eq:Gsingle}
\end{equation}
Here, the geometrical continuum-model parameters $R_0$ and $\delta$ contribute only to a shift (first term) of the free energy, whereas all essential adsorption information is incorporated in the second term.

In \Fig~\ref{fig:single}c we compare the transfer free energies $\Delta G$ in the collapsed PNIPAM (empty symbols) from this study with the adsorption free energies $G_\trm{ads}$ on a single chain (full symbols) obtained from $\Gamma'$ \chgA{for the elongation $\lambda=0.83$}   (Ref.~\citen{kanduc2017selective}) using \Eq~\ref{eq:Gsingle}. As can be seen, similarly to $\Delta G$, also $G_\trm{ads}$ scales approximately linearly with the surface area of the solute, albeit with a different slope.
A linear fit of \Eq~\ref{eq:Ga2} to the nonpolar data points for $G_\trm{ads}$ gives the value $\gamma_\trm{m}^\trm{(ads)}=-9(2)$~kJ\,mol$^{-1}$\,nm$^{-2}$. 
That is, the molecular surface tension of the adsorption on a single chain is roughly half the value of the transfer of nonpolar solutes into the collapsed state ($\gamma_\trm{m}=-18(1)$~kJ\,mol$^{-1}$\,nm$^{-2}$).
Neglecting the offsets of the fits, we can deduce an empirical relation
\begin{equation}
G_\trm{ads}\approx \Delta G/2
\label{eq:Gads}
\end{equation}
Its simple interpretation can be that as a solute adsorbs on a single chain it exposes only a half of its surface to the chain, whereas the other half faces the water phase.
It is also interesting that the adsorption free energies are almost insensitive to the polarity of the solutes:  As has been shown~\cite{kanduc2017selective}, a polar solute that adsorbs on the chain points its OH group away from the chain, thus making it less relevant for the adsorption. In contrast, if the solute solvates inside the collapsed PNIPAM phase, the OH group has to be accommodated within other polymer chains, making its contribution much more significant.

\chgA{
Note that the PNIPAM models differ a bit in the two studies. 
PNIPAM chains in Ref.~\citen{kanduc2017selective} are isotactic, whereas they are atactic in this study. As has been shown~\cite{kanduc2017selective}, the adsorption varies by less than 25$\%$ among different tacticities, which yields only around 1~kJ/mol variance in the adsorption free energy (\Eq~\ref{eq:Gsingle}).
 The other difference is that} the force field for PNIPAM in Ref.~\citen{kanduc2017selective} is the standard OPLS-AA~\cite{opls1988}, which slightly differs in partial charges from the one used in this study, yet we believe that this should not alter our qualitative conclusions.

\subsection{Partitioning: Swollen vs.\ collapsed state}
We now make a step further and use the observation given by \Eq~\ref{eq:Gads} to relate the partition \chgA{ratios} in the swollen and collapsed state of a hydrogel.
Assuming that the gel in the swollen state is very dilute, such that adjacent polymer chains are elongated and far apart, and that the effects of cross-linkers can be ignored, we can calculate the total number of solutes inside the gel, $N_\trm{in}$, from the adsorption on all its chains,
\begin{equation}
\chgA{N_\trm{in}=c^\trm{(out)} V+ \Gamma' \lambda L_\trm c c^\trm{(out)}}
\end{equation}
The first term corresponds to the non-adsorbed solute background in the gel and the second term is the number of adsorbed solutes on all the chains with a total contour length $L_\trm{c}$. 
\chgA{Using \Eq~\ref{eq:Gsingle} and} expressing the contour length in terms of the polymer volume fraction, $\phi_\trm{p}=\pi {R_0}^2L_\trm c/V$, we arrive at the following expression for the partition \chgA{ratio} in the swollen state
\begin{equation}
K^\trm{(s)}=1+\phi_\trm{p} (2\chgA{\lambda}\delta/R_0)\,\rme^{-G_\trm{ads}/\kB T}
\end{equation}
Evaluating the numerical factor $2\chgA{\lambda}\delta/R_0\simeq 1$ \chgA{(note that $\lambda\approx 0.8$ is close to unity for a very swollen gel)} and using \Eq~\ref{eq:Gads} together with \Eq~\ref{eq:K}, we finally obtain the relation between the partition \chgA{ratios} for nonpolar solutes in the collapsed ($K$) and swollen ($K^\trm{(s)}$) states
\begin{equation}
K^\trm{(s)}\simeq 1+\phi_\trm{p}^\trm{(s)}\sqrt{K}
\label{eq:KKs}
\end{equation}
In our model of the collapsed state of a PNIPAM gel, the partition \chgA{ratio} for larger aromatic molecules is  $K\sim10^3$--$10^4$ (\Fig~\ref{fig:K}b). Now, assuming a typical polymer volume fraction of  $\phi_\trm{p}=0.1$ in the swollen state, the above equation gives the the partition \chgA{ratio} in the swollen state of only $K^\trm{(s)}\simeq4$--$10$, which we also obtained in our previous work~\cite{kanduc2017selective}.
This simple relation demonstrates an enormous impact that a changing environment from hydrophilic (swollen state) to hydrophobic (collapsed state) has on the partitioning.	
Namely, doubling the hydrophobic interaction in the collapsed state as suggested by \Eq~\ref{eq:Gads}, results in a quadratic relation of the partition \chgA{ratios}, $K\propto {K^\trm{(s)}}^2$.

On the other hand, polar solutes are subject to more specific interactions, such that we cannot design a similar simple rule. We found for instance that polar aromatic molecules follow similar trend as nonpolar molecules, thus \Eq~\ref{eq:KKs} is applicable for them as well. On the other hand, for the two alcohols one can rather consider $G_\trm{ads}\approx\Delta G$, which results in a linear scaling of both partition \chgA{ratios}, namely $K^\trm{(s)}\simeq 1+\phi_\trm{p}^\trm{(s)}K$.

\section{Conclusions}

We have performed extensive all-atom molecular dynamics simulations of solvation of small subnanometer-sized solutes in collapsed PNIPAM polymers in equilibrium with water. 
Sorbed and heterogeneously distributed water between the polymer chains, whose content depends on temperature, causes non-uniform partitioning of the solutes.
Nonpolar solutes are preferentially expelled from water clusters and tend to reside in `dryer' regions of the gel, whereas polar molecules tend to partition closer to or in water clusters.

The transfer free energy from water into PNIPAM of nonpolar solutes scales very well with an effective molecular surface area, thus making it the decisive descriptor. The presence of a hydroxyl group in generally opposes the solvation in PNIPAM, but its nature is not as deterministic as the surface area, especially for aromatic molecules.
The partition \chgA{ratios} of the studied small solute molecules, spanning from neon to benzene derivatives, span over five orders in magnitude.

Finally, an important observation that we find is a strong correlation between the transfer free energies into a collapsed PNIPAM and the adsorption free energies to a single PNIPAM chain. The adsorption free energies are about one half the values of the transfer free energies. This can be explained in a sense that an adsorbed solute exposes only a half of its surface to the polymer chain and the other half faces the water phase, whereas a solvated solute exposes its entire surface area to the PNIPAM phase. As a consequence, the partition \chgA{ratio} in the collapsed state scales quadratically with the partition ratio in the swollen state. That means that the release/uptake of the gel undergoing a swelling/collapse is particularly large for larger nonpolar molecules.

 Understanding the solvation mechanisms inside not only PNIPAM but also other responsive hydrogels is important for the rational design of novel materials. The next important step in understanding these systems would be to analyze the solvation properties of ions and the effects of cross-linkers in these gel models, which are our future objectives.


\subsection*{Conflict of Interest} The authors declare no competing
financial interest.

\subsection*{Supporting Information Description}
Solvation in water: MD vs.\ experiments; Solute--solute contribution;
 Sensitivity to the effective solute surface area

\subsection*{Acknowledgments}
The authors thank Yan Lu and Matthias Ballauff for useful discussions.
 This project has received funding from the European Research Council (ERC) under the European Union's Horizon 2020 research and innovation programme (Grant Agreement No.\ 646659-NANOREACTOR). M.K.~acknowledges the financial support from the Slovenian Research Agency (research core funding no.\ P1-0055).
The simulations were performed with resources provided by the North-German Supercomputing Alliance~(HLRN).

\setlength{\bibsep}{0pt}
\bibliography{literature_revision}

\end{document}